# Mechanical properties of freely suspended atomically thin dielectric layers of mica

Andres Castellanos-Gomez[1,2,*], Menno Poot[1,3], Albert Amor-Amorós[2], Gary A. Steele[1], Herre S.J. van der Zant[1], Nicolás Agraït[2,4], and Gabino Rubio-Bollinger[2]

[1] Kavli Institute of Nanoscience, Delft University of Technology, Lorentzweg 1, 2628 CJ Delft, The Netherlands

[2] Departamento de Física de la Materia Condensada, Universidad Autónoma de Madrid, Campus de Cantoblanco, E-28049 Madrid, Spain

[3] Department of Engineering Science, Yale University, Becton 215, 15 Prospect St., New Haven, CT 06520, USA

[4] Instituto Madrileño de Estudios Avanzados en Nanociencia IMDEA-Nanociencia, E-28049 Madrid, Spain

a.castellanosgomez@tudelft.nl

**ABSTRACT**

We have studied the elastic deformation of freely suspended atomically thin sheets of muscovite mica, a widely used electrical insulator in its bulk form. Using an atomic force microscope, we carried out bending test experiments to determine the Young's modulus and the initial pre-tension of mica nanosheets with thicknesses ranging from 14 layers down to just one bilayer. We found that their Young's modulus is high (190 GPa), in agreement with the bulk value, which indicates that the exfoliation procedure employed to fabricate these nanolayers does not introduce a noticeable amount of defects. Additionally, ultrathin mica shows low pre-strain and can withstand reversible deformations up to tens of nanometers without breaking. The low pre-tension and high Young's modulus and breaking force found in these ultrathin mica layers demonstrates their prospective use as a complement for graphene in applications requiring flexible insulating materials or as reinforcement in nanocomposites.

## 1. Introduction

Graphene-like atomically thin crystals [1] have emerged as powerful candidates for future flexible electronic applications such as touch screens, smart textiles or wearable biosensors. Graphene has been already employed as transparent and flexible conducting electrodes, showing outstanding mobility and mechanical properties [2–4]. Future flexible electronic applications will need, however, not only thin flexible conductors but also semiconductors and insulators. This fact has motivated recent work on the fabrication and characterization of atomically thin semiconductors, such as molybdenum disulphide ($MoS_2$) [5–9], and insulators, such as boron nitride [10–13] or muscovite mica [14–18]. While thin semiconductors can be used in flexible field effect transistors, atomically thin insulators can be used as substrates [19–21], dielectrics or tunnel barriers [22, 23]. Despite the recent interest in ultrathin boron nitride substrates [10–13], commercially available boron nitride single crystals are typically smaller than 45 μm in size which hampers their application in large scale devices. Muscovite mica, on the other hand, is already employed in several areas of electronics due to its low cost, flexibility, light transparency, high dielectric strength and high chemical/thermal stability. Additionally, muscovite mica can be modified on demand with high precision by atomic force microscope-induced wear [17, 25]. A recent report by He et al. [24] demonstrated the fabrication of large area organic field-effect transistors based on a mica single crystal gate insulator (100 nm in thickness). Although this work suggests the prospective use of ultrathin mica crystals as dielectric in flexible organic devices, the mechanical properties of this ultrathin insulator have not yet been reported.

Here, we present a study of the mechanical properties of freely suspended nanosheets of muscovite mica. Flakes with thicknesses ranging from 14 layers down to just 2 layers have been subjected to a bending test experiment carried out with an atomic force microscope (AFM) [8]. Our measurements allow us to obtain the pre-tension, the Young's modulus and the breaking force of these atomically thin crystals, all of which are crucial parameters for their application in flexible electronics.

## 2. Experimental

### 2.1 Sample fabrication

Atomically thin mica flakes were directly deposited by mechanical exfoliation [15, 26] onto a 285 nm $SiO_2$/Si substrate pre-patterned with holes 1–1.1 μm in diameter and 200 nm in depth in order to fabricate freely suspended mica sheets without exposing them to a lithographic resist or an electron beam which can contaminate or introduce defects in the crystal lattice [27, 28]. Polydimethylsiloxane (PDMS) stamps [15, 29, 30], commonly used in soft-lithography [31, 32], were used for flake deposition instead of the commonly used Scotch tape to avoid leaving adhesive traces on the surface which could contaminate the fabricated flakes [33].

### 2.2 Optical microscopy

Optical microscopy was used to identify ultrathin mica flakes. Flakes thinner than 20 nm can be easily identified by illuminating with 600 nm light because of their strong negative contrast (Fig. 1) [15]. The optical identification of ultrathin mica flakes was carried out with a Nikon Eclipse LV100 microscope under normal incidence with a 50x objective (0.8 numerical aperture). The optical micrographs were acquired with a Canon EOS 550D digital camera attached to the optical microscope. Quantitative optical microscopy measurements (right inset in Fig. 1) were carried out using narrow-bandpass filters (purchased at Edmund Optics) to select the illumination wavelength.

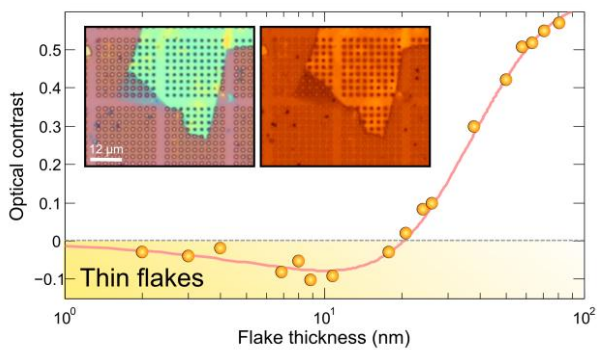

**Figure 1** Optical contrast as a function of the thickness measured for mica flakes deposited on top of a 285 nm $SiO_2$/Si substrate with an illumination wavelength of 600 nm. Flakes thinner than 20 nm present a negative optical contrast, facilitating their identification. The solid line is the calculated optical contrast using a Fresnel law-based model, assuming the refractive index of the flakes to be equal to that of bulk mica. (Inset, left) Optical microscopy image of a mica flake with two regions with different number of layers. (Inset, right) The optical image, acquired with an illumination wavelength of 600 nm, showing that there is a region with less than 20 layers in thickness given its negative contrast.

### 2.3 Atomic force microscopy

Once a thin flake had been identified, a combination of quantitative optical microscopy [15] and AFM was used to accurately determine the thickness of the mica flake (Fig. 2). The AFM characterization of the flakes was carried out with a Cervantes AFM (from Nanotec Electrónica) operated at room temperature and under ambient conditions. The flake thickness was determined by contact mode AFM to avoid artifacts in the thickness measurements [34]. The number of layers was determined by dividing the measured thickness by the interlayer distance in mica (1 nm). The suspended area of the nanosheets was carefully characterized by AFM in order to exclude suspended sheets with wrinkles or a varying number of layers from the subsequent study of their mechanical properties.

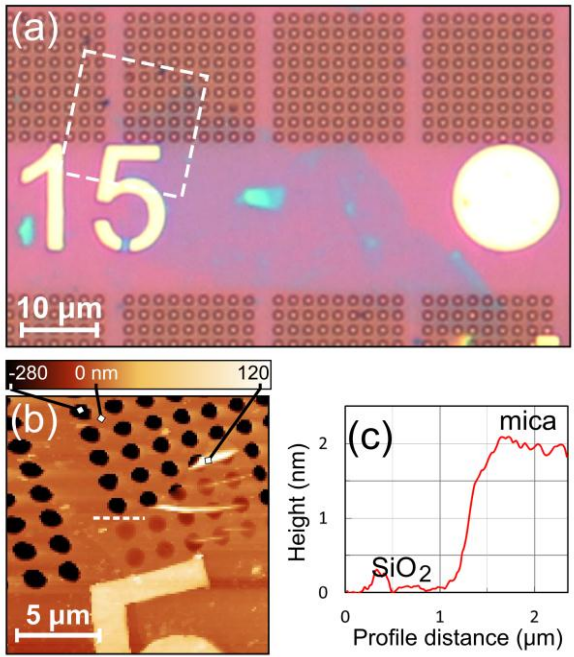

**Figure 2** (a) Optical microscopy image of a bilayer mica flake deposited on top of a 285 nm SiO$_2$/Si substrate pre-patterned with an array of holes 1–1.1 μm in diameter (b) Contact mode AFM topography acquired on the region marked by a dashed rectangle in (a). (c) Topographic line profile acquired along the dashed line in (b).

It should be noted that Raman spectroscopy, commonly employed to determine the number of layer in graphene, does not allow the determination of the number of layers in atomically thin mica flakes [15]. As reported previously [15], although single layer flakes can be identified by optical microscopy and AFM, they are usually not much larger than 1 × 1 μm² (and therefore do not cover any hole) while thin multilayer mica flakes (2 to 20 layers) can be larger than 10 × 10 μm² (forming several suspended mica sheets out of the same flake).

**2.4 Force versus displacement traces**

AFM was used to study the elastic deformations of the suspended mica layers in a nanoscopic version of a bending test experiment (see inset in Fig. 3) [8]. The tip of the AFM was employed to apply a central point load on the suspended nanosheets while the deflection of the cantilever was measured. In order to accurately position the tip in the center of the suspended part of the nanosheets several AFM scans were carried out, decreasing the scansize to avoid effects due to the non-linearity and the creep of the piezoelectric scanner. The overlap between consecutive AFM images was checked to determine whether or not there was thermal drift (and eventually to correct for this drift). We found that it is possible to locate the tip in the center of the suspended nanosheets with less than 30 nm of uncertainty. Note that an uncertainty of ± $R/10$ in the determination of the center of the suspended nanosheet does not produce a significant modification, within the experimental resolution, of the force versus deflection characteristic traces [35].

The relationship between the deflection of the cantilever ($\Delta z_c$) and the deformation of the flake ($\delta$) is given by

$$\delta = \Delta z_{piezo} - \Delta z_c \qquad (1)$$

where $\Delta z_{piezo}$ is the displacement of the AFM piezoelectric tube during the force load cycle. Using Eq. (1), the deformation of the nanolayers was obtained from force versus displacement traces. The AFM cantilever deflection $\Delta z_c$ was calibrated by measuring force versus displacement traces on the SiO$_2$/Si substrate. Once the AFM tip is in contact with the hard substrate, a displacement of the AFM scanning piezo $\Delta z_{piezo}$ (which itself is calibrated by measuring the height of monoatomic steps of graphite) produces a deflection of the cantilever ($\Delta z_c$) equal to $\Delta z_{piezo}$. We have additionally checked that the deformation produced by the AFM tip in atomically thin mica sheets on the SiO$_2$/Si substrate (non-suspended) is negligible for the typical force values employed in this manuscript.

## 3. Results and discussion

Figure 3 shows three representative deformation versus force ($F(\delta)$ hereafter) curves, measured for nanosheets 2, 6 and 12 layers in thickness. Note that negative forces up to –40 nN can be applied to the suspended nanosheets due to the tip–mica adhesion forces. For small deflections, these $F(\delta)$ traces are linear and their slope defines the effective spring constant of the nanolayer ($k_{eff}$) [36]

$$k_{eff} = \left.\frac{\partial F}{\partial \delta}\right|_{\delta=0} = \frac{4\pi E}{3(1-v^2)} \cdot \left(\frac{t^3}{R^2}\right) + \pi T \quad (2)$$

where $E$ is the Young's modulus, $v$ is Poisson's ratio ($v$ = 0.25 for mica [37]), $t$ is the thickness, $R$ is the radius of the suspended nanosheet and $T$ is the initial pre-tension of the nanosheet. The first term in expression (2) corresponds to the spring constant of a plate with a certain bending rigidity. The second term represents the spring constant of a stretched membrane. From the slope of a $F(\delta)$ trace alone it is not possible to individually determine both the Young's modulus and the pre-tension because they are entangled in the effective spring constant.

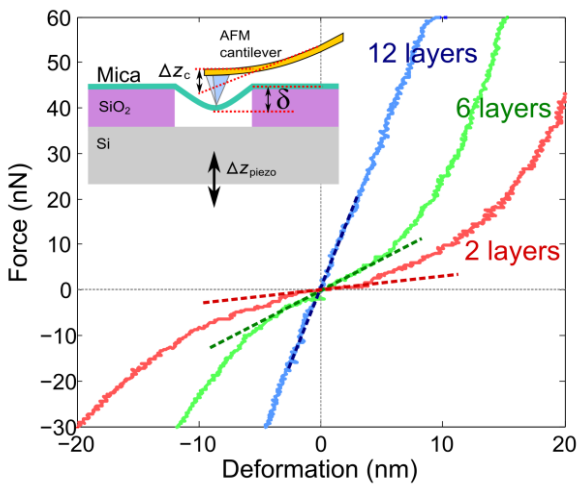

**Figure 3** Force versus deformation traces measured at the center of the suspended part of mica nanosheets 2, 6 and 12 layers in thickness. The slope of the traces around zero deflection is marked by a dotted line. (Inset) schematic diagram of the bending test experiment carried out on a freely suspended mica nanosheet.

A technique to independently determine $E$ and $T$ is to measure the spatial profile of the effective spring constant, instead of measuring the $F(\delta)$ at the center only [38]. Alternatively, the thickness scaling of Eq. (2) can be used; the first term of the spring constant (which accounts for the bending rigidity of the nanosheet) depends on the sheet thickness while the second one (the pre-tension) is assumed to be independent of thickness. Therefore, to determine $E$ and $T$ independently one can measure the effective spring constant for different sheet thicknesses and fit the experimental data to Eq. (2), assuming $E$ and $T$ to be independent of thickness. Figure 4 shows the measured $k_{eff}$ as a function of the thickness of 27 different mica layers. The solid line corresponds to the $k_{eff}$ as a function of $t$ calculated with Eq. (2) using $E$ = 170 ± 40 GPa and $T$ = 0.15 ± 0.15

N/m. The light red area around the fit indicates the uncertainty on the fit.

An alternative procedure to determine the Young's modulus and the pre-tension of these suspended mica layers relies on the study of the non-linear dependence of the $F(\delta)$ traces measured at the center of the sheet. Mica nanosheets thinner than 7 layers show strongly non-linear $F(\delta)$ traces, while for thicker sheets the $F(\delta)$ traces are linear (see Fig. 3). This thickness dependence results from an interesting mechanical property of these suspended nanolayers: with decreasing thickness, there is a crossover from a bending-dominated (i.e. plate) to a stretching-dominated (i.e. membrane) behavior [39]. The relationship between the applied force at the center of the flake and the resulting deformation is [36, 39, 40]

$$F = \left[\frac{4\pi E}{3(1-\nu^2)} \cdot \left(\frac{t^3}{R^2}\right)\right]\delta + (\pi T)\delta + \left(\frac{q^3 E t}{R^2}\right)\delta^3 \qquad (3)$$

where $q = 1/(1.05 - 0.15\nu - 0.16\nu^2)$ is a dimensionless parameter. Expression (3) shows that apart from the two linear terms, coming from the flake spring constant in Eq. (2), the relationship between the force and the deformation presents a cubic term that takes into account the stiffening of the layer during the force load cycle which makes $F(\delta)$ nonlinear. Note that for ultrathin sheets this expression is not strictly correct: here a correction to the bending rigidity term (the first term) should be introduced to account for the discrete number of layers [38]. We have found, however, that for ultrathin layers (less than 5 layers), the bending rigidity term is negligible in comparison with the deformation-induced tension term (the third term), and thus this correction is effectively smaller than 1%. Note that the effect of shearing stresses, on planes parallel to the surface of the plate, can be neglected because of the low sheet thickness to sheet radius ratio. For the mica layers studied here, this ratio is $t/R < 0.02$ and thus the correction due to this effect is also below 1%.

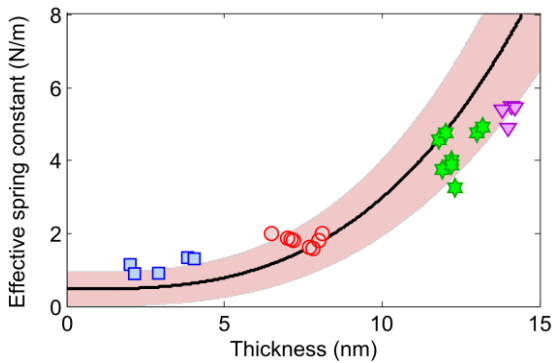

**Figure 4** Effective spring constant measured for 27 suspended mica nanosheets as a function of their thickness. Data points sharing color and symbol correspond to suspended nanosheets from the same mica flake. The experimental data have been fitted to Eq. (2) using $E = 170$ GPa and $T = 0.15$ N/m (black solid line) with an uncertainty of $\Delta E = \pm 40$ GPa and $\Delta T = \pm 0.15$ N/m (gray area around the black line).

Both the Young's modulus and the initial pre-tension of an individual suspended mica sheet can be obtained by fitting non-linear $F(\delta)$ traces to Eq. (3). Figure 5(a) shows the fit of a representative $F(\delta)$ to Eq. (3) for a trilayer mica nanosheet from which one obtains $E = 200 \pm 30$ GPa and $T = 0.20 \pm 0.03$ N/m. We have additionally found that bilayer mica flakes can stand reversible deformations up to tens of nanometers (see Fig. 3) but they tend to break for deformations around 30–40 nm (with a maximum force load $F_{max} = 60$–150 nN). One can roughly estimate the corresponding breaking force $\sigma_{max}$ of the nanolayers by using the expression for the indentation of an elastic membrane by a spherical indenter [39, 40]

$$\sigma_{max} = \sqrt{\frac{F_{max} E}{4\pi r_{tip} t}} \qquad (4)$$

If we assume a tip radius $r_{tip}$ of 15–20 nm (according to the manufacturer), the estimated breaking force of the mica bilayers is $\sigma_{max}$ = 4–9 GPa), which is in agreement with the microindentation hardness of bulk muscovite mica [41]. Due to the high breaking force value of atomically thin mica layers, some 2–5% of the Young's modulus value, they are good candidates for flexible electronics applications requiring a crystalline dielectric substrate. Note that other dielectric materials commonly used in electronics (such as silicon oxide) can only withstand deformations up to 0.06 % of their Young's modulus [42].

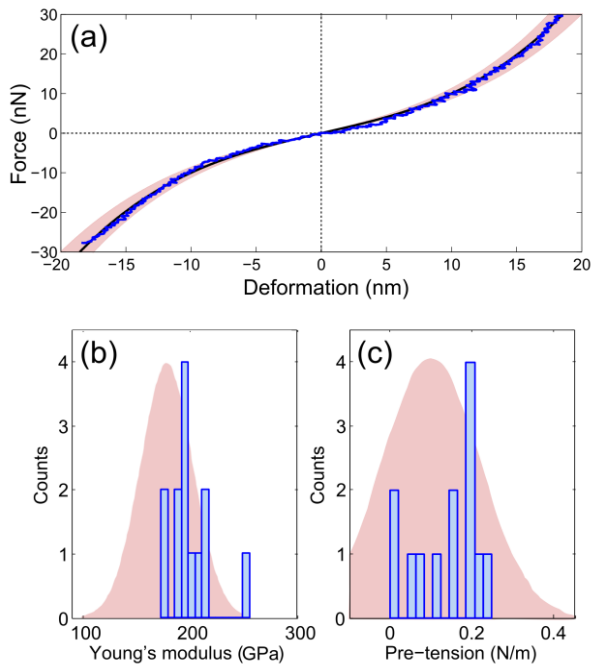

**Figure 5** (a) Non-linear force versus deformation measured on a suspended trilayer mica flake. The experimental data have been fitted to Eq. (3), with $E$ = 200 GPa and $T$ = 0.2 N/m (solid black line). The light red area around the fit indicates the uncertainty of the fit with $\Delta E$ = ±30 GPa and $\Delta T$ = ±0.025 N/m. (b) and (c) Histogram of the Young's modulus and the initial pre-tension obtained by fitting $F(\delta)$ curves (measured for 13 sheets 2 to 8 layers thick) to Eq. (3). The light red Gaussian peaks in panels (b) and (c) indicate the mean and standard deviation of the Young's modulus and the pre-tension obtained from the analysis of the effective spring constant for different thicknesses (see Fig. 3) to facilitate the comparison.

We have carried out a statistical analysis of the flake-to-flake variation in $E$ and $T$, which can be due to different densities of defects in the flakes and/or adhesion force with the substrate. Figures 5(b) and 5(c) show the histogram of the Young's modulus and the pre-tension obtained for 13 sheets 2 to 8 layers thick by fitting their force versus. deformation traces to Eq. (3). The mean values for the Young's modulus and pre-tension and their standard deviation are $E$ = 202 ± 22 GPa and $T$ = 0.14 ± 0.08 N/m. These values are in good agreement with those obtained by analysis of the thickness dependence of the effective spring constant (Fig. 3). To facilitate this comparison, Figs. 5(b) and 5(c) include two Gaussian peaks that indicate the mean and standard deviation of $E$ and $T$ values obtained from Fig. 3. We did not observe any clear evidence of thickness dependence of the tension and Young's modulus of the 13 flakes studied (with thicknesses ranging from 2 nm to 8 nm). This supports the assumption of thickness-independent pre-tension made in the analysis of Fig. 4. The maximum pre-tension observed in the mica nanosheets (<0.25 N/m) is lower than that observed in similar nanosheets (few-layer graphene flakes fabricated by mechanical exfoliation show pre-tensions up to 1 N/m), which is desirable for the fabrication of highly tunable nanomechanical systems.

It is important to note that for some nanomaterials with a high surface-to-volume ratio, such as ZnS nanobelts, the elastic modulus can be 50% lower than the bulk counterpart due to the lower constraint of surface atoms (thereby making these nanomaterials easier to deform in the elastic regime than their bulk form) [43]. This degradation of mechanical properties resulting from the reduction of the material dimensions is not observed in ultrathin mica, making mica nanolayers appropriate for mechanical applications requiring a high surface-to-volume ratio, such as fillers in nanocomposites. Indeed, despite the high surface-to-volume ratio of atomically thin mica sheets, their Young's modulus is comparable to the value of 176.5 ± 1.1 GPa determined by Brillouin scattering in much thicker crystalline mica samples [44]. It is also important to compare the Young's modulus values of ultrathin mica layers (~190 GPa) with those of other two-dimensional (2D) insulators such as graphene oxide (200 GPa) [45], or hexagonal boron nitride (250 GPa) [13], carbon nanosheets (10–50 GPa) [46] and 2D clays (22 GPa) [47]. Atomically thin mica is also competitive with commonly used thin insulating films (grown by atomic layer deposition) such as $Al_2O_3$ or $HfO_2$ (220 GPa) [48]. Therefore, atomically thin mica flakes are a promising material for highly demanding mechanical applications requiring ultrathin dielectrics.

## 4. Conclusions

We have studied the mechanical properties of freely suspended mica nanosheets 2 to 14 layers thick with a bending test experiment performed with an atomic force microscope. The Young's modulus of these suspended nanosheets is high, $E$ = 202 ± 22 GPa, making them competitive with other insulating 2D crystals such as boron nitride or thin films of oxides such as $HfO_2$. We have also found that these suspended nanolayers present low pre-strain (<0.25 N/m) and high breaking force (4–9 GPa). The mechanical properties of atomically thin mica crystals (i.e. low pre-tension and high Young's modulus and breaking force) validate their applicability in highly demanding mechanical applications such as flexible ultrathin insulating substrates/dielectrics or for reinforcement in nanocomposites.

## Acknowledgements

This work was supported by MICINN (Spain) through the programs MAT2008-01735, MAT2011-25046 and CONSOLIDER-INGENIO-2010 'Nanociencia Molecular' CSD-2007-00010, Comunidad de Madrid through program Nanobiomagnet S2009/MAT-1726 and the European Union (FP7) through the program RODIN.